\documentclass{article}
\usepackage[utf8]{inputenc}
\usepackage{float}
\usepackage{color}
\usepackage{graphicx}
\usepackage{amsmath}
\usepackage{authblk}

\usepackage{graphicx}

\graphicspath{{figures/}}
\usepackage{geometry}
\title{\textbf{Widefield lensless endoscopy via speckle-correlations}}

\author[1,2]{Amir Porat}
\author[3]{Esben Ravn Andresen}
\author[3]{Hervé Rigneault}
\author[4]{Dan Oron}
\author[2]{\\ Sylvain Gigan}
\author[*,1,2]{Ori Katz}

\affil[1]{\small Department of Applied Physics, The Selim and Rachel Benin School of Computer Science \& Engineering, The Hebrew University of Jerusalem, Jerusalem 9190401, Israel}
\affil[2]{Laboratoire Kastler Brossel, Université Pierre et Marie Curie, Ecole Normale Supérieure, CNRS, Collège de France, 24 rue Lhomond, 75005 Paris, France}
\affil[3]{Aix-Marseille Université, CNRS, Centrale Marseille, Institut Fresnel, UMR 7249, Domaine Universitaire de Saint Jérôme, F-13397 Marseille Cedex 20, France}
\affil[4]{Department of Physics of Complex Systems, Weizmann Institute of Science, Rehovot 76100, Israel \vspace{5mm}}

\affil[*]{ Author to whom correspondence and requests for materials should be addressed: orik@mail.huji.ac.il}

\date{}

\begin{document}

\maketitle

\par

\textbf{Flexible fiber-optic endoscopes provide a minimally-invasive solution for imaging at depths beyond the reach of conventional microscopes. Current endoscopes require focusing and/or scanning mechanisms at the distal end, which limit miniaturization and frame-rate, and induce aberrations. Alternative lensless solutions are based on adaptive wavefront-correction, but are extremely sensitive to fiber bending.
Here, we demonstrate a novel endoscopic approach, which enables single-shot imaging at a variable working distance through a conventional fiber bundle, without the use of any distal optics. Our approach computationally retrieves the object image from a single speckle pattern transmitted through the bundle, exploiting phase information preserved through inherent angular speckle correlations.
Unlike conventional fiber-bundle endoscopes, the resulting image is unpixelated, the resolution is diffraction-limited, objects can be imaged at variable working distance, and the technique is completely insensitive to fiber bending. Since no optical elements are required, miniaturization is limited only by the bundle diameter.
}

\section*{}
Flexible optical endoscopes are an important tool in biomedical investigations and clinical diagnostics. They enable imaging at depths where scattering prevents noninvasive microscopic investigation. An ideal microendoscopic probe should be flexible, allow real-time diffraction-limited imaging at various working distances from its facet, while maintaining a minimal cross-sectional footprint \cite{Flusberg,oh}.

Single-mode fibers (SMF) can be used as small diameter light-guides for endoscopic imaging, but in order to obtain two-dimensional (2D) images a mechanical scanning head \cite{Flusberg,oh} or a spectral disperser \cite{syl,mertz} should be mounted at the distal end of a fiber, sacrificing frame-rate and probe size or resolution.
Alternatively, 2D image information can be delivered by the different modes of a multimode fiber (MMF), if the complex phase randomization and mode mixing of the MMF is measured and compensated for, computationally or via wavefront-shaping \cite{mosk, holo, holo2, cizmar2, papa, choi, cizmar, shamir}. However, the extreme sensitivity of the wavefront-correction to any movement or bending of the fiber necessitates direct access to the distal end for recalibration, or precise knowledge of the bent shape \cite{cizmar}.

A robust and widely-used type of imaging endoscopes is based on fiber bundles, constructed from thousands of individual cores packed together, each carries one image pixel information. Imaging is performed in a straightforward manner if the target object is positioned immediately adjacent to the bundle's facet (Fig.1a) \cite{Flusberg,nenad}. While straightforward to implement, conventional fiber bundle endoscopes suffer from limited resolution and pixelation artifacts dictated by the individual cores and cladding diameters, and from a fixed working distance, which locates the imaging plane directly at the bundle's facet unless distal optics are added. When the object is placed away from this fixed image plane, only a blurred, seemingly information-less image appears at the proximal facet (Fig.1b).

The fundamental reason for the limitation to a fixed imaging plane is that spatial phase information is scrambled upon propagation through the bundle, due to the different random core-to-core phase delays. Although these phase distortions can be measured and compensated for using a spatial light modulator (SLM) \cite{french,Herve,herve2,choi2,PsaltisMemEff}, the sensitivity of the phase correction to fiber bending severely limits applicability, in a similar manner to the case of MMF. As a result, conventional bundle imaging techniques work under the assumption that phase information is lost and thus rely on intensity-only information transmission by each core.

Despite these seemingly fundamental restrictions, here we show that important spatial phase information is retained in the speckle patterns produced by propagation through any fiber bundle. We demonstrate that this information can be used to overcome both the fixed working distance and the bend-sensitivity limitations of current approaches. Moreover, we demonstrate that all that is required to utilize this information is to computationally analyze a single image of the speckle pattern transmitted through the fiber.
We thus present a simple approach that performs widefield endoscopic imaging at an arbitrary working-distance from a bare fiber bundle, using only a conventional camera, without any phase correction or pre-calibration, or any distal optics.
Our single-shot, diffraction-limited, and pixelation-free imaging technique is based on exploiting inherent angular speckle correlations, and is inspired by the recent advancements in imaging through opaque scattering barriers \cite{kb,OOO, taka}, and by methods used to overcome atmospheric turbulence in astronomy \cite{Labeyrie}).

\subsection*{Principle}

\begin{figure}[H]
\centering
\includegraphics[width=\textwidth,height=\textheight,keepaspectratio]{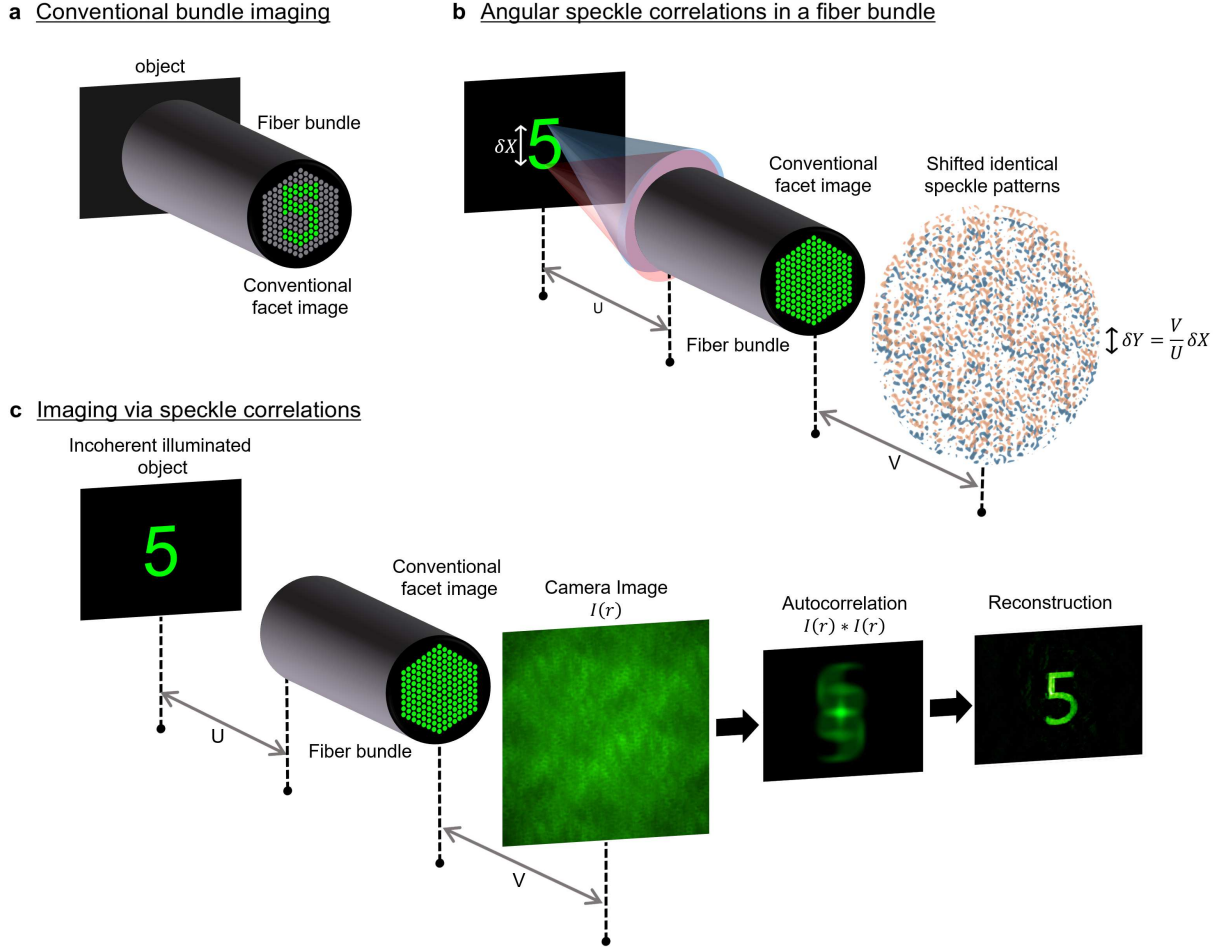}
\caption{
Conventional vs. speckle-correlations based fiber bundle endoscopy (numerical results): \textbf{(a)}, In a conventional fiber bundle endoscope, the intensity image of an object placed adjacent to the input facet is transferred to the output facet. \textbf{(b)}, When the object is placed at a distance, $U$, from the input facet, only a blurred information-less image is formed at the output facet. However, even though two different point sources at the object plane (red and blue) produce similar patterns at the output facet, the relative tilt between their wavefronts at the bundle input results in a proportional spatial shift in the speckle pattern observed at a small distance, $V$, from bundle's output facet. \textbf{(c)}, For an extended object, the image of the light intensity at the distance $V$ is the sum of many shifted speckle patterns, and its autocorrelation provides an estimate to the object's autocorrelation. The diffraction-limited object image is retrieved from this autocorrelation via a phase-retrieval algorithm.
}
\label{fig:sketch}
\end{figure}

The underlying principle for our technique is presented in (Fig.\ref{fig:sketch}). The propagation of light through an imaging fiber bundle composed of single-mode cores is characterized by the fact that each core, $i$, in the bundle preserves the intensity of the light coupled to it, but randomizes the transmitted phase by adding a different phase in each core, $d\phi_i$.
Therefore, a point source that is placed at a distance $U$ from the bundle input facet (the object plane), will produce a speckle pattern at a distance $V$ from the bundle output facet, due to the added random phase pattern to the otherwise spherical wavefront (Fig.1b). A second point source placed at the same object plane, but shifted in transverse position by a distance $\delta X$ relative to the first point source, will produce a nearly identical speckle pattern at the image plane, but shifted by $\delta Y=\frac{V}{U} \cdot \delta X$, due to the angular tilt of $\delta X / U$ of the input wavefront (Fig.1b). Thus, within the angular range in which the two speckle patterns are highly correlated, they present a shift-invariant point spread function (PSF) of the fiber bundle. This angular range is analogous to the isoplanatic patch in adaptive optics \cite{adap}, and to the angular 'memory-effect' for speckle correlations in scattering media \cite{freund, mem_prl}. For an ideal fiber bundle, with randomly positioned single-mode cores and no core-to-core coupling, the angular correlation range is essentially the core's numerical aperture (NA) (See derivation and experimental verification in supplementary information section 3).

As a direct result, when a spatially incoherently illuminated object is contained within this angular range, the light from every point on the object forms correlated, but shifted, speckle patterns at a distance from the output facet (Fig.\ref{fig:sketch}c). The image of the light at this plane will be the  intensity sum of these identical shifted speckle patterns. Building on the recent results in imaging through opaque barriers \cite{OOO}, the image of the object itself can be computationally recovered from the autocorrelation of this image (Fig.\ref{fig:sketch}c). The mathematical justification for this is straightforward: due to the angular correlations, the image of the light intensity measured far enough from the output facet can be described by a simple convolution between the object's intensity pattern $O(r)$, and the single (unknown) speckle pattern $PSF(r)$:
\begin{equation} \label{eq1}
I(r)=O(r)*PSF(r)
\end{equation} \\
Taking the autocorrelation of this intensity image $I(r)$ gives:
\begin{equation} \label{eq2}
I(r) \star I(r) = (O(r) \star O(r)) * (PSF(r) \star PSF(r))
\end{equation}

Since the autocorrelation of a random speckle pattern $PSF(r) \star PSF(r)$ is a sharply-peaked function having a peak with a width of a diffraction limited spot, the autocorrelation of the raw image, $I(r) \star I(r)$,  will approximate the autocorrelation of the object itself (up to a statistical average over the number of captured speckle grains and a constant background term \cite{OOO}, see discussion).
Thus, the autocorrelation of a single camera image of the light propagated through the bundle is essentially identical to the target object's autocorrelation, and one can directly reconstruct the original object from this autocorrelation using a phase retrieval algorithm \cite{fienup,kb,OOO},Fig.\ref{fig:sketch}c.\\

\section*{Results}
\subsection*{Single-shot imaging via speckle correlations}

\begin{figure}[H]
\centering
\includegraphics[width=0.6\textwidth,keepaspectratio]{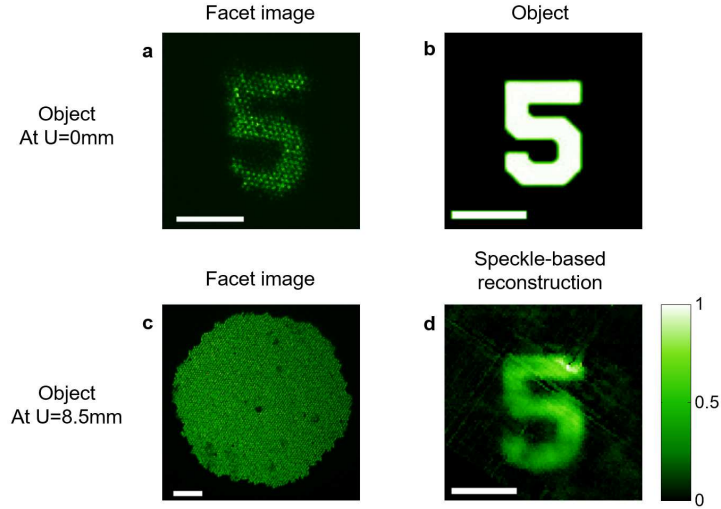}
\caption{Experimental demonstration and comparison to conventional bundle imaging. \textbf{(a)}, conventional imaging through a fiber bundle requires the object to be placed at U=0mm. \textbf{(b)}, The original object. \textbf{(c)}, conventional imaging through a fiber bundle when the object is placed at U=8.5mm from the bundle's input facet. No imaging information is obtainable. \textbf{(d)}, Same situation as in (c) but using the presented speckle-based approach; All scalebars are $100\mu m$.}
\label{fig:sketch_res}
\end{figure}

\begin{figure}[H]
\centering
\includegraphics[width=0.75\textwidth,keepaspectratio]{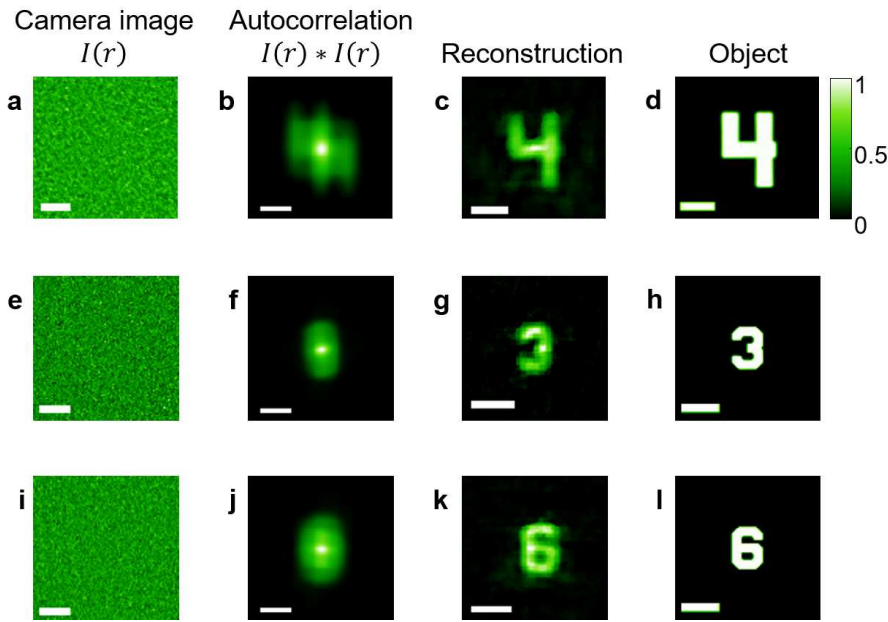}
\caption{Experimental single-shot imaging of various objects at different working distances, $U$: \textbf{(a)}, Raw camera image. \textbf{(b)}, Autocorrelation of (a). \textbf{(c)}, Object reconstruction from (b). \textbf{(d)}, object image. \textbf{(e-l)}, same as (a-d) for different objects from 1951 USAF target. In (a-h) $U=161mm$, in (i-l) $U=65mm$. Scalebars: (a,e)=$10mm$, (i)=$5mm$, (b,c,d,f,g,h)=$0.5mm$, (j,k,l)=$0.25mm$.}
\label{fig:rec}
\end{figure}

To test the speckle-based single-shot approach we have performed several proof-of-concept experiments, whose results are presented in Fig.\ref{fig:sketch_res} and Fig.\ref{fig:rec}, using the setup depicted in Fig.\ref{fig:sketch}c. In these experiments a target object illuminated by a spatially incoherent narrowband source was placed at various distances from a 530$\mu m$ diameter fiber bundle having 4500 cores (see Methods and Supplementary Figure 1). The image of the object was reconstructed from a single image of the light pattern measured at a small distance from the bundle proximal end.

Fig.\ref{fig:sketch_res} gives a comparison between our technique and conventional imaging through a fiber bundle. In this geometry, since the object is located at a distance from the bundle's distal facet, no information is obtainable in the conventional approach (Fig.\ref{fig:sketch_res}.c), whereas in our technique the object's image is retrieved with diffraction limited resolution (Fig.\ref{fig:sketch_res}.d). Several additional experimental examples are presented in Fig.\ref{fig:rec}, which gives the raw camera speckle images, their autocorrelations, and the images reconstructed from them, side-by-side with the original objects.
The technique is not restricted to transmission geometry, and works equally well in reflection geometry where the light source is placed adjacent to the bundle end (and in principle can be provided by the fiber itself), as is demonstrated in supplementary figure 2.

\begin{figure}[H]
\centering
\includegraphics[width=\textwidth,height=\textheight,keepaspectratio]{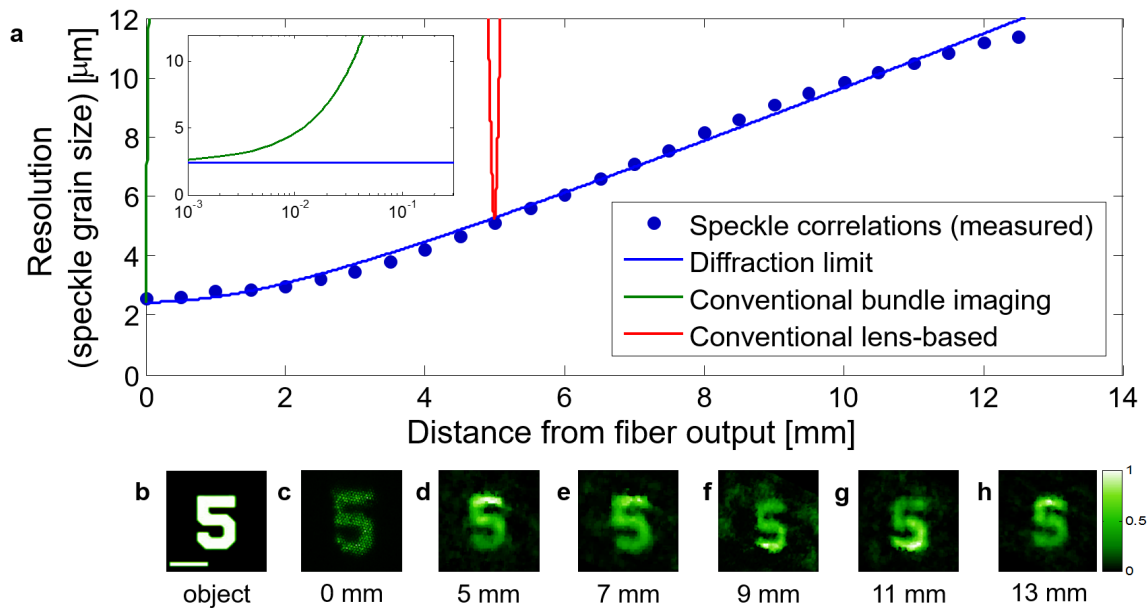}
\caption{Imaging resolution at various working distance. \textbf{(a)}, Resolution (speckle grain size) as a function of the object's distance from the bundle facet ($U$): Measured (blue circles); theoretical diffraction limit according to Eq.\ref{eq3}) with  $D_{bundle}=570\mu m$ and $d_{mode}=2.4\mu m$ (blue line). Calculated resolution of conventional lensless bundle imaging (green line); Calculated resolution of conventional lens-based bundle imaging assuming a distal objective with a focus at $U=5mm$ (red line).  (b). \textbf{(b)}, Object from the 1951 USAF target, group 3, Scalebar=$100\mu m$. \textbf{(c)}, Conventional bundle imaging of (b) when placed at $U=0$, \textbf{(d-h)}, Speckle-based single-shot reconstructions of (b) at distances of 5-13mm. }
\label{fig:z}
\end{figure}

The resolution of this speckle-based technique is dictated by the speckle grain dimensions, which are diffraction-limited\cite{OOO}, and thus give a resolution that would be obtained by an ideal aberration-free optical system, having the same diameter aperture \cite{Labeyrie}). 
In Fig.\ref{fig:z} we experimentally characterize this imaging resolution as a function of the object's distance from the bundle's facet ($U$) (see Methods), and compare it to the resolution of a conventional bundle based endoscope with and without distal optics.
The speckle grain size $\delta x$ (and the resolution) follows approximately the formula:
\begin{equation} \label{eq3}
\delta x \approx \sqrt{(\frac{\lambda}{D_{bundle}} \cdot U)^2 + d_{mode}^2}
\end{equation}
where $\lambda$ is the wavelength, $D_{bundle}$ is the fiber bundle's diameter, $U$ is the distance between the object and the fiber facet, and $d_{mode}$ is the mode field diameter of a single core.
The diffraction limited resolution that is attained at any distance from the fiber provides a very large range of working distances, as we demonstrate by imaging the digit 5 of group 3 from the USAF target at various distances in Fig.3d-h.

\subsection*{Applicability to broadband and coherent light}

\begin{figure}[H]
\centering
\includegraphics[width=0.5\textwidth,keepaspectratio]{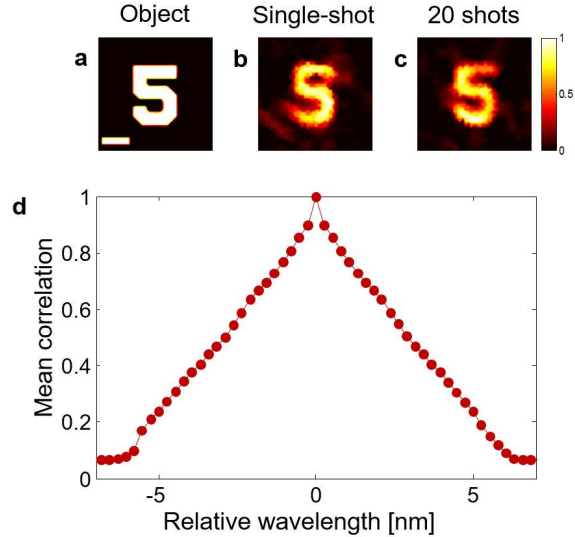}
\caption{Speckle imaging with broadband illumination. \textbf{(a)}, Object pattern. \textbf{(b)}, Reconstruction of (a) from a single speckle image through the bundle with illumination of 10nm bandwidth (800nm central wavelength). \textbf{(c)}, Increased reconstruction fidelity via autocorrelation ensemble averaging over 20 different camera shots. Each shot provides uncorrelated speckle realization by slight bending of the the bundle. \textbf{(d)}, Experimentally measured spectral correlations of the 48.5cm long fiber bundle used (Schott 153333385); Scalebar=$100\mu m$.}
\label{fig:spec}
\end{figure}

To consider the applicability of the approach to broadband illumination and as a step towards fluorescence imaging, the experiments of Fig.2-3 were repeated with a broadband illumination light source (central wavelength 800nm, spectral width 10nm). Their results are presented in Fig.\ref{fig:spec}. Broadband illumination can be used without appreciably affecting the performance of the technique, as long as the illumination bandwidth is narrower than the fiber bundle's speckle spectral correlation bandwidth \cite{mosk,cao}, since the speckle patterns produced by different wavelength within this bandwidth stay well correlated. This spectral bandwidth is Fourier transform related to the time delay spread of the light propagating in the different cores. Fig.\ref{fig:spec}.d presents the characterization of the spectral correlation width of the imaging bundle used in this experiment. The broadest spectral correlation width is attained for a bundle having SMF cores (rather then MMF ones \cite{cao}). Working with illumination bandwidth that is larger than the spectral correlation width is possible, but will reduce the camera image contrast \cite{OOO}.

In the case of spatially coherent illumination, one can follow a similar derivation as in (Eq.1-2) for objects which are placed at the far field of the bundle, by replacing all of the light intensity terms with their complex field amplitudes (see derivation in Supplementary Information section 6).
Interestingly, instead of having to measure the complex speckle field and calculate its autocorrelation, one can simply use the intensity image of the bundle's facet, which is related to this coherent autocorrelation by a Fourier transform. Thus, the complex coherently-illuminated object can be reconstructed via phase-retrieval from a single image of the bundle's facet intensity, which is the object diffraction pattern, in a manner analogous to x-ray coherent diffraction imaging \cite{xray}. Supplementary Fig.6 provides a simple experimental proof-of-principle for this approach. Objects which are placed closer to the facet may be reconstructed via Fresnel phase retrieval \cite{fresnel}.

\section*{Discussion}

We have presented a widefield microendoscopic technique that offers diffraction-limited imaging resolution at a variable working distance without the use of any distal optics. The simple and calibration-free technique is straightforward to implement as it utilizes essentially the same setup already used in a conventional fiber bundle endoscope. All that is required is to shift the camera imaging plane away from the bundle proximal facet. Unlike conventional microendoscopes, our technique is aberration free, it does not suffer from pixelation artifacts. Compared to novel approaches that are based on active correction of the fiber wavefront distortions \cite{papa, cizmar, french, herve2}) our technique is insensitive to fiber bending and works inherently with spatially incoherent illumination in a single-shot, without the need for scanning.

Computationally retrieving an image from a speckle field, rather than  conventional direct imaging, dramatically alters the influence of the bundle parameters on the imaging performance. The major differences are that: (1) Local defects that are naturally present in some bundle cores and conventionally lead to local spatial information loss, translate in our technique only to a reduced number of speckles. This affects only the autocorrelation spatial averaging, and thus distributes the local loss of information, leading to a small decrease in the autocorrelation signal to noise (SNR) (see Fig.\ref{fig:sketch_res}a,d); (2) Any crosstalk between neighboring cores, which conventionally reduces resolution and contrast, will only affect the FOV by reducing the speckle angular correlation width (see derivation in Supplementary Information section 3); (3) The diameter of each core and the spacing between the cores, which conventionally limit the imaging resolution and induce pixelation artifacts, do not directly affect the resolution in the speckle based approach; (4) The FOV in our technique is not fixed by the bundle outer diameter but is dictated by the fiber NA at large working distances, and can be scaled at will by changing the working distance (see discussion below). Interestingly, since the FOV is given by the NA of the cores (for SMF cores), all of the light that is guided by the bundle cores can be used for imaging. This is in contrast to the case of speckle-based imaging through highly scattering media \cite{OOO}, where the FOV narrows down with increasing barrier thickness, limiting the technique in many cases to small, isolated objects; (5) The number of cores, which conventionally affects only the number of resolution-cells, dictates  also the number of speckles in a single image that can be used for calculating the autocorrelation function. A too low number of speckles can lead to insufficient ensemble averaging, which in turn reduces the signal to noise ratio (SNR) of the autocorrelation. This is especially important when imaging large objects whose angular dimensions are comparable to the bundle's NA, since the larger spatial coordinates in the autocorrelation have less spatial averaging. This potential difficulty can be overcome by averaging the autocorrelation over multiple shots of different speckle patterns  (as is done in stellar speckle interferometry \cite{Labeyrie}). Different uncorrelated speckle patterns can be easily produced in many ways. The simplest approach is to slightly move or bend the bundle, as we demonstrate experimentally in Fig.\ref{fig:spec}. Alternative approaches for ensemble averaging include using orthogonal polarizations (supplementary figure 4), or different spectral bands (Fig.5d). Still, even with a single shot and only a few thousands cores, we have demonstrated that the spatial averaging is sufficient to perform single-shot imaging (Figs.2-3)).

Similar to the recently introduced computational endoscopic and spectroscopic techniques \cite{syl,mertz,cao}, the object image is computationally extracted from a relatively low contrast camera image. The raw image contrast is the result of the incoherent sum of shifted speckle patterns, and is inversely related to the square-root of the number of bright resolution cells on the object \cite{OOO}. This contrast level dictates the required camera well depth and integration time \cite{OOO,Dainty}.\\

As demonstrated in Fig.4, our approach allows the imaging of objects placed at a large range of working distances. However, for optimal performance the working distance needs to be large enough to ensure that the light from each point on the object is collected by all of the bundle's cores (as in wavefront-shaping based approaches \cite{french,Herve,herve2}). This optimal minimum working distance, $U_{min}$ is given by: $U_{min}=\frac{D_{bundle} \cdot d_{mode}}{\lambda} = \frac{D_{bundle}}{NA}$, where $\lambda$ is the wavelength, $D_{bundle}$ is the total diameter of the bundle, $d_{mode}$ is the mode field diameter of a single core, and NA is the a single core's numerical aperture. $U_{min}$ can be reduced by using a smaller diameter bundle with higher NA cores \cite{NA} (which will also increase the FOV), or by adding a thin scattering layer at the distal end. If a shorter working-distance is desired one may simply splice a glass rod spacer to the distal end. Another possible alternative is to acquire speckle images from sub-apertures of the fiber, which will decrease $U_{min}$ at the expense of resolution. Interestingly, the information contained in several sub-apertures speckle images may be used to retrieve depth information \cite{taka}. Axial sectioning and increased resolution may also be possible by the use of structured illumination \cite{cathie,nenad}. \\

Additional possible extensions include the use of temporal gating to provide efficient axial-sectioning capability \cite{Beaurepaire}, and improved computational retrieval by exploiting phase information contained in the image bi-spectrum \cite{Dainty}, or in speckle images captured under different aperture masks. While dealing with very different length scales, additional approaches used to overcome atmospheric turbulence in astronomical observations should be applicable to bundle-based endoscopy, since the challenge is essentially the same: imaging through an aperture with an unknown random phase mask \cite{Dainty}.

Besides these possible improvements the proposed speckle correlation approach has the advantage of being extremely simple to implement, requires no distal optics, no wavefront control, and immediately extend the imaging capabilities of any fiber bundle based endoscopes. Quite uniquely, and contrary to many
novel endoscopic imaging techniques, fiber movements are not a hurdle but are beneficial, and are exploited rather than fought against to provide better imaging quality.

\section*{Methods}
\textbf{Experimental set-up.} The complete experimental set-ups for incoherent and coherent imaging are presented in Supplementary Information section 1.
The fiber bundles used were two Schott fiber bundles. One, which was used for the experiments of Figs.2-4, had 4.5k cores with 7.5 $\mu$m inter-core distance, 0.53 mm diameter, and a length of 105 cm. The second, which was used for the coherent and broadband experiments, had 18k cores with 8 $\mu$m inter-core distance, a diameter of 1.1 mm, and a length of 48.5 cm (Schott part number: 15333385).
The imaged objects were taken from a USAF resolution target (Thorlabs R3L3S1N).
In the experiments of Figs.2-4, the objects were illuminated by a narrow bandwidth spatially incoherent pseudothermal source at a wavelength of 532 nm, based on a Coherent Compass 215M-50 cw laser and a rotating diffuser (see Supplementary Information section 1). In the coherent experiments the same laser was used without a rotating diffuser. In the experiment of Fig.5 a Ti:Sapphire laser with a bandwidth of ~12nm around a central wavelength of 800nm (Spectra-physics Mai Tai) and a rotating diffuser was used.
The camera used in experiments of Figs.2-4 was a PCOedge 5.5 (2,560x2,160 pixels). The camera integration time was 10 milliseconds to 2 seconds (typically a few hundred milliseconds).
The objects were placed at distances of 5-250 mm from the bundle's input facet and the camera was placed at distances of 5-50 mm from the bundle's output facet, or behind an objective that imaged the light close to the bundle's output facet.
For the experiment of Fig.5, the objects were placed at distances of 45-70 mm from the bundle's input facet, and the camera was placed at distances of 5-20 mm from the bundle's output facet.

\par
\textbf{Image processing.} For the incoherent images, the raw camera image was spatially normalized for the slowly varying envelope of the transmitted light pattern by dividing the raw camera image by a low-pass-filtered version of it that estimated its envelope.
The autocorrelation of the processed image was calculated by an inverse Fourier transform of its energy spectrum (effective periodic boundary conditions).
The resulting autocorrelation was cropped to a rectangular window with dimensions ranging between 40x40 pixels and 400x400 pixels (depending on the imaged object dimensions), and the minimum pixel brightness in this window was background-subtracted from the entire autocorrelation trace. In addition, the intensity of the central pixel of the autocorrelation was taken as equal to one of its neighbors, to reduce the effect of camera noise.
A two-dimensional Tukey window applied on the autocorrelation was found to enhance the phase-retrieval reconstruction fidelity in some of the experiments.
For the coherent case, the raw image intensity was thresholded to remove background noise and the image was zero-padded before calculating its Fourier transform.

\par
\textbf{Phase-retrieval algorithm.} The phase-retrieval algorithm was implemented according to the recipe provided by Bertolotti and co-authors \cite{kb} (for details see Supplementary section 5). The object constraints used were it being real and nonnegative, or belonging to only half of the complex plane, with a positive real part. The algorithms were implemented in Matlab. The  reconstructed images were median-filtered, and the images of (Fig.\ref{fig:spec}) were also 2D interpolated.


\par
\section*{Acknowledgements}
This work was supported by the LabEx ENS-ICFP: ANR-10-LABX-0010/ANR-10-IDEX-0001-02 PSL; the CNRS/Weizmann NaBi European Associated Laboratory;  The Aix-Marseille University/ Hebrew University of Jerusalem collaborative-research joint program. The work was funded by the European Research Council (grant no. 278025). O.K. was supported by the Marie Curie Intra-European fellowship for career development (IEF).
\\
The authors cordially thank Mickael Mounaix and Cathie Ventalon for their valuable help.

\section*{Author contributions}
D.O. proposed the idea. A.P. and O.K. performed the theoretical analysis. A.P. performed the experiments, numerical work and results analysis. All authors contributed to designing the experiments and writing the manuscript. S.G. and O.K. supervised the project.

\end{document}